 \documentstyle[aps,prc,twocolumn,epsfig,floats]{revtex}
 \draft
\newcommand{\etal}{\emph{et al.}}
\begin{document}
\renewcommand{\thefootnote}{\arabic{footnote}}
\twocolumn[\columnwidth\textwidth\csname@twocolumnfalse\endcsname
\title{Comment on ``Structure of exotic nuclei and superheavy elements\\
       in a relativistic shell model''}
\author{Michael Bender}
\address{Service de Physique Nucl{\'e}aire Th{\'e}orique,
         Universit{\'e} Libre de Bruxelles,
         CP 229, 
         B-1050 Bruxelles, Belgium}
\date{29 July 2002}
\maketitle
%
%
\begin{abstract}
A recent paper (M. Rashdan, Phys. Rev. C \textbf{63}, 044303 (2001)) 
introduces the new parameterization NL-RA1 of the relativistic 
mean-field model together with a new parameterization of the 
constant gap pairing model. Some conclusions of the paper
may be doubtful as the pairing model is unrealistic and the 
known ground-state deformation of particular nuclei is neglected.
\end{abstract}
\addvspace{3mm}]
\narrowtext
%
%
A recent study \cite{Ras01a} introduces the parameterization 
NL-RA1 of the relativistic mean-field model (RMF). It is left open 
which data NL-RA1 is fitted to, which prevents to relate its 
properties to the fitting strategy. NL-RA1 is 
compared with early parameterizations as NL1 or NL-SH, 
more recent ones as NL3 or NL-Z2 are not considered.
Extrapolation of NL-RA1 to superheavy nuclei contradicts 
earlier studies. As will be shown here, conclusions drawn 
in \cite{Ras01a} may be doubtful as the pairing model used
is unrealistic, and nuclei known to be deformed are
calculated assuming spherical shapes.
%
%
\begin{figure}[b!]
\centerline{\epsfig{file=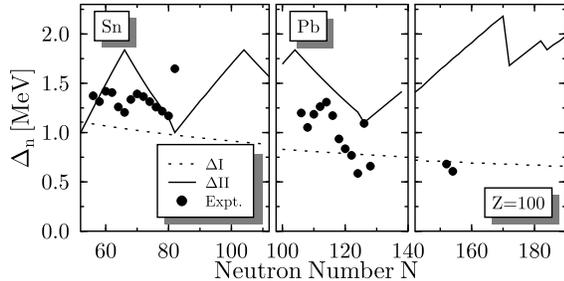,width=7.5cm}}
\caption{\label{fig:gaps}
Neutron pairing gaps $\Delta_{n}$ from models $\Delta$I and $\Delta$II 
in comparison with with experimental five-point gaps $\Delta^{(5)}$. 
The peak of $\Delta^{(5)}$ at closed shells is spurious, see
\protect\cite{Ben00c}. The pattern of the gaps from model $\Delta$II 
around \mbox{$N \approx 180$} is caused by the fact that the
parameters $N_{c1}$ and $N_{c2}$ of model $\Delta$II do not 
correspond to the nearest neutron shell closures.
}
\end{figure}
%
%

Ref.~\cite{Ras01a} employs two parameterizations of the constant-gap 
model, one of which ($\Delta$I) has been used in many early applications 
of the RMF. The pairing matrix elements are independent on the 
single-particle levels which is unrealistic for loosely-bound systems 
\cite{Dob84a} as those discussed in \cite{Ras01a}. The pairing gap, 
related to the odd-even mass staggering, has to be parameterized 
as a function of $N$ and $Z$ \cite{Mol92b}. Model $\Delta$I
describes the average behaviour of the pairing gap. Introducing 
about 30 parameters $N_{c1}$ and $N_{c2}$, model $\Delta$II 
attempts to incorporate the reduction of the pairing gap around 
known magic numbers, but remains arbitrary for exotic systems where 
shells might be quenched or new shells occur. Model $\Delta$II misses 
the overall reduction of the pairing gap with $A$, see Fig.~\ref{fig:gaps}. 
For transactinides, gaps are overestimated by a factor of 3.
(There is currently much discussion about blocking and 
mean-field contributions to calculated pairing gaps that are neglected
here, see \cite{Ben00c,Dug02b} and references therein. Those corrections 
cannot be easily incorporated into the simplistic constant gap
model. Their contribution is usually smaller than 20 $\%$ 
and decreases with $A$.)
There is no justification for model $\Delta$II as it fails by construction 
to describe the size of the pairing gap, which is the key observable 
for pairing correlations. As the pairing gap determines the occupation of 
the single-particle states around the Fermi surface, most results 
presented in \cite{Ras01a} are affected in one way or the other.

%
%
\begin{figure}[t!]
\centerline{\epsfig{file=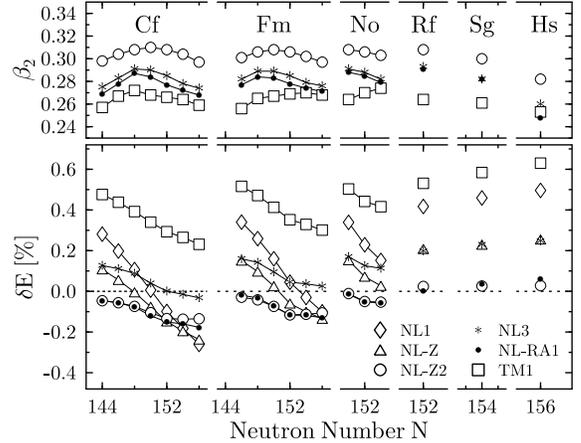,width=7.5cm}}
\caption{\label{fig:sh}
Ground-state quadrupole deforma\-tion $\beta_2 =  4 \pi \langle r^2 \, Y_{2 0} 
\rangle/(3 A r_0^2)$ with $r_0 = 1.2 \, A^{1/3}$ fm
and relative error on the binding energy $\delta E$
of even-even trans\-acti\-nide nuclei. See \protect\cite{Bue98a}
for a discussion of the uncertainty of $\delta E$.
}
\end{figure}
%
%

In \cite{Ras01a}, the predictive power of NL-RA1 for superheavy nuclei 
is tested for the heaviest known even-even nuclei as earlier done in 
\cite{Bue98a}. 
Results in \cite{Ras01a} differ significantly from \cite{Bue98a}.
One reason are different pairing models, but there is a second one.
In \cite{Ras01a} it is not mentioned which shape degrees
of freedom are accounted for. Repeating the calculations indicates
that all results in \cite{Ras01a} are obtained assuming 
spherical shapes. This is consistent with experiment for 
Sn and Pb isotopes, but there is agreement among 
all successful mean-field models that the known superheavy nuclei
are deformed. This is confirmed by experiment for selected 
isotopes up to $^{254}$No \cite{Reiter}. In \cite{Ras01a} the 
missing deformation energy (which is on the order of 10 MeV or 0.5$\%$) 
is replaced  by the artificially increased pairing correlation energy 
from model $\Delta$II.

Using a more realistic state-dependent delta pairing force with 
parameters adjusted along the strategy of \cite{Ben00c}
and allowing for deformation change significantly the systematics 
of $\delta E$ for transactinide nuclei, see Fig.\ \ref{fig:sh}. 
Comparing with Fig.\ 17 in \cite{Ras01a}, all forces perform better. 
Similar changes can be expected for the values given in Fig.\ 18 of 
\cite{Ras01a} (see \cite{Cwi99a} for complications 
when calculating odd-$A$ nuclei which are neglected in \cite{Ras01a}.)
The change in $\delta E$ when comparing NL1 with NL-Z and NL-Z2 reflects 
an improved center-of-mass correction \cite{Ben00b} and the inclusion 
of data on exotic nuclei into an otherwise identical fit.
NL-RA1 and NL-Z2, however, have the same good quality for 
binding energies of transactinide nuclei in spite their very different 
nuclear matter properties. 
%
%
\begin{figure}[t!]
\centerline{\epsfig{file=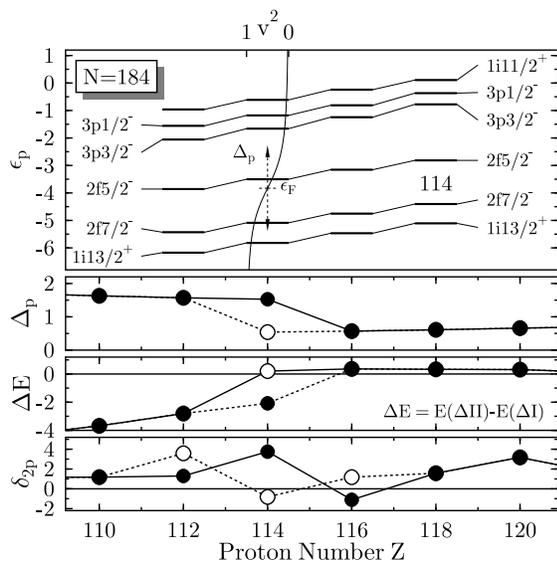,width=7.5cm}}
\caption{\label{fig:d2p}
Single-particle spectrum of the protons, pairing gap $\Delta_p$ 
in model $\Delta$II, difference $\Delta E$ between the binding energy 
obtained with model $\Delta$I and $\Delta$II, and two-proton gap 
$\delta_{2p}$ (all in MeV) for \mbox{$N=184$} isotones. The uppermost 
panel also displays the occupation probability $v_k^2$ for the protons 
in $^{298}114$ calculated with model $\Delta$II. The arrows denote 
the pairing gap $\Delta_p$, $\epsilon_p$ the Fermi energy. Results 
for $^{298}114$ depend sensitively on the value of $\Delta_p$.
Filled (open) markers denote calculations
where $\Delta_p$ for $^{298}114$ is calculated using the prescription 
for nuclei with $N$ smaller (larger) than \mbox{$Z=114$}. (Ref.\
\protect\cite{Ras01a} leaves it open which one to use. Results 
presented there correspond to filled markers).
}
\end{figure}
%
%

In the framework of mean-field models, a magic number is associated with 
a large gap in the single-particle spectrum which causes a 
discontinuity in the systematics of binding energies. Those (and other) 
discontinuities are filtered from data with the two-proton gap
$\delta_{2p}(N,Z) = E(N,Z-2) - 2 E(N,Z) + E(N,Z+2)$ and the similar 
quantity for neutrons. $\delta_{2p}$ has to be taken with care, as
it assumes the structure of the considered nuclei does not change,
which is not necessarily fulfilled for heavy nuclei \cite{Ben02a}.
Ground-state deformation of some of the nuclei might quench $\delta_{2p}$
and has to be considered as soon as one wants to predict future data 
on these nuclei. To demonstrate 
the non-existence of the spherical \mbox{$Z=114$} shell, however, 
spherical calculations are sufficient and enforce the validity of 
$\delta_{2p}$ as a signature for magicity.
Figure \ref{fig:d2p} displays the key quantities that reveal the origin 
of the large $\delta_{2p}$ for $^{298}$114 found in \cite{Ras01a}.
Let us look at filled markers first.
Single-particle energies in $^{298}$114 do not change significantly 
when varying the pairing strength. The pairing gap for $^{298}$114
from model $\Delta$II (\mbox{$\Delta_p = 1.7$} MeV) is of the same 
order as the \mbox{$Z=114$} gap in the single-particle spectrum (1.4 MeV). 
About 3 protons occupy levels above the \mbox{$Z=114$} gap, 
inconsistent with the assumption of a major shell closure.
By construction, $\Delta_{p}$ drops by a factor 2 at \mbox{$Z=114$}. 
This causes a discontinuity in the pairing correlation energy which 
is clearly visible when comparing binding energies obtained 
with models $\Delta$I and $\Delta$II.

The discontinuity in the pairing correlation energy built into model 
$\Delta$II causes the large value for $\delta_{2p}$($^{298}$114) 
found in \cite{Ras01a}, not the underlying shell structure.
This is confirmed when moving the discontinuity of $\Delta_p$ to 
the non-magic proton number \mbox{$Z=112$}, see the open markers 
in Fig.~\ref{fig:d2p}. $\delta_{2p}$ is now peaked at $^{296}$112, 
which has no closed spherical proton shell. $\delta_{2p}$ cannot be 
used as a signature for shell 
closures when as the pairing gap and its fluctuations are of similar size 
as the spacing of single-particle energies. However, the size 
of the pairing gap in model $\Delta$II is unrealistic anyway. 
Calculations with more realistic pairing models do not show any 
significance for a major shell closure at \mbox{$Z=114$}, consistent 
with \cite{Rut97a}.

The \mbox{$Z=120$} shell is not considered in the choice of $N_{c1}$ 
and  $N_{c2}$ in pairing model $\Delta$II, the pairing gap 
has huge mid-shell values there. This smears out the 
\mbox{$Z=120$} shell effect found in \cite{Rut97a} in terms of 
$\delta_{2p}$ (c.f.\ Fig.~21 in \cite{Ras01a}) but again does not 
affect the single-particle spectra.
%
%

\end{document}